\documentclass[parskip=full]{scrartcl}
\usepackage{apacite}
\usepackage{graphicx}
\usepackage{longtable}
\usepackage[utf8]{inputenc}
\usepackage{booktabs}
\usepackage{url}

\title{The Gig's Up: How ChatGPT Stacks Up Against Quora on Gig Economy Insights}
\author{Thomas Lancaster\thanks{This paper is primarily intended as a demonstration of how ChatGPT can be used to enhance research studies, so it should not be considered as a fully formed, checked and polished research paper. A motivated researcher with prompting knowledge and a prepared template could produce papers of this style and standard on a daily basis.}}
\date{February 2024}

\begin{document}

\maketitle

\section*{Abstract}
Generative AI is changing the way in which humans seek to find answers to questions in different fields including on the gig economy and labour markets, but there is limited information available about closely ChatGPT simulated output matches that obtainable from existing question and answer platforms. This paper uses ChatGPT as a research assistant to explore how far ChatGPT can replicate Quora question and answers, using data from the gig economy as an indicative case study. The results from content analysis suggest that Quora is likely to be asked questions from users looking to make money and answers are likely to include personal experiences and examples. ChatGPT simulated versions are less personal and more concept-based, including considerations on employment implications and labour rights. It appears therefore that generative AI simulates only part of what a human would want in their answers relating to the gig economy. The paper proposes that a similar comparative methodology would also be useful across other research fields to help in establishing the best real world uses of generative AI. 

\section{Introduction}

This paper brings together interest in how generative AI tools such as ChatGPT may be replacing traditional human labour, with a timely comparison as to how closely ChatGPT can simulate written human answers to question related to the gig economy. Production of this paper has been substantially aided by having access to ChatGPT as a form of research assistant.

The inspiration for this paper has been a substantial body of quickly published work comparing human and ChatGPT responses to medical questions. The study reported here considers as an alternative the evolution of the human workforce. This uses the question and answer platform Quora, which is perhaps indicative of the gig economy in action, since individuals who successfully answer questions can be paid for their contributions and the number of views received. If question and answer sites are replaced by generative AI written answers, this may therefore have an effect on those income streams. A discussion may need to had if the readership and use of such sites would be impacted.

The gig economy, characterised by its flexible job opportunities and reliance on freelance contracts, has been impacted by generative AI. Studies such as \citeauthor{jin_2023_generate} \citeyear{jin_2023_generate} and \citeauthor{demirci_2023_who} \citeyear{demirci_2023_who} have shown a marked decrease in demand for tasks that AI can replicate, highlighting the disruptive potential of generative AI on traditional freelance jobs. Conversely, platforms like Quora, which facilitate knowledge sharing and expertise exchange, play a crucial role in the digital information ecosystem, although the ways in which Quora.com can be misused by students looking to breach academic integrity have also been noted \cite{lancaster_2022_the}. The work of \citeauthor{wang_2013_wisdom} \citeyear{wang_2013_wisdom} underscores the importance of understanding the drivers of growth and engagement on such platforms, as they are said to contribute to the collective intelligence of the internet.

This paper aims to use content analysis to explore the intersection of AI technologies with the gig economy and question and answer sites, examining the diversity of their views and impact on knowledge about this field. In doing so, this study draws upon a diverse body of literature, including the transformative potential of ChatGPT in the gig economy outlined by \citeauthor{huang_2023_chatgpt} \citeyear{huang_2023_chatgpt}, the empirical analysis of AI's impact on job posts by \citeauthor{demirci_2023_who} \citeyear{demirci_2023_who}, and the strategic adaptation to AI technologies discussed by \citeauthor{yilmaz_2023_aidriven} \citeyear{yilmaz_2023_aidriven}. Furthermore, this is led by the broader implications of AI integration into the workforce as surveyed by \citeauthor{george_2023_chatgpt} \citeyear{george_2023_chatgpt} and the metaphorical analysis of the platform economy by \citeauthor{vallas_2020_what} \citeyear{vallas_2020_what}. Through this comparative analysis, this paper aims to contribute to the scholarly discussion on the understanding of work in a world of generative AI, how this is perceived in both human and generative AI led question and answer sites, and the evolving nature of these platforms.

\section{Methods}
A simple systematic process was used to collect data for this investigation. ChatGPT (running GPT 4) was used to support the data capture process, as well as the subsequent analysis.

To collect results from Quora, an incognito search of Google.com was undertaken for the phrase "gig economy" with results restricted to Quora. Text was extracted from the first 10 results viewable without needing to log into a Quora account. The question and a ChatGPT summarised version of the top answer were saved.

Inside a fresh ChatGPT session, the tool was asked to itself act as a simulation of Quora. It was asked to return what it believed the first 10 results of a Quora search for "gig economy" would be, including the question and a summarised version of the answer. This would give data comparable with the previous search.

The results from both data collection activities were compared using a ChatGPT aided process of content analysis. This was to identify categories of interest and to investigate how sensible it would be to replace Quora with ChatGPT as a research source.

\section{Results}

Table \ref{quora-table} shows the first 10 questions obtained from a Google search for "gig economy" restricted to Quora, along with a ChatGPT summarised version of the answer given.

\begin{longtable}{*{11}{|p{5.6cm}}|}%{|1|1|}
\hline \multicolumn{1}{|c|}{\textbf{Question}} & \multicolumn{1}{c|}{\textbf{Summary of Answer}} \\ \hline 
\endhead
\endfoot

What is the gig economy, and how can one participate in it? & The gig economy or access economy is where ownership is temporary and you rent everything, from human resources to lawn mowers, facilitated by aggregators like Uber, AirBnB, and Freelancers. It creates creative destruction of industries but offers lower prices and requires individuals to act as companies. \\
\hline
How does the gig economy impact the way people earn money in today's workforce? & The gig economy can offer new lifestyle choices and opportunities outside of traditional employment, though it may not pay as well as jobs in good corporations. It allows for flexibility but comes with challenges like irregular income and competition on platforms like Upwork. \\
\hline
Have you explored the gig economy or sharing economy for income opportunities? & Yes, the gig economy offers income opportunities through short-term or freelance work, facilitated by online platforms. It has become popular for its flexibility but presents challenges such as irregular income and limited benefits. \\
\hline
What are the pros and cons of pursuing a career in the gig economy? & Pros include flexibility, diverse opportunities, potential for competitive wages, entrepreneurial experience, and better work-life balance. \\
\hline
What are some profitable opportunities in the gig economy in 2024? & Profitable opportunities depend on offering skills to the world, leveraging personal networks, and taking advantage of the flexibility to create a competitive marketplace presence. \\
\hline
What is the gig economy? What is the digital economy? & The gig economy is characterized by short-term, flexible jobs, often facilitated by digital platforms, whereas the digital economy encompasses economic activities conducted through digital technologies, including e-commerce and digital content creation. \\
\hline
What impact has the gig economy had on recent events? & The gig economy allows for reduced costs and increased flexibility for businesses by hiring workers on a temporary or contract basis, affecting traditional business models. \\
\hline
How can the gig economy be regulated to protect workers' rights and job security? & Addressing the gig economy's transient nature and preparing for technological advancements like robotaxis and drones are crucial. Current legislation has struggled to protect gig-economy workers effectively. \\
\hline
The 'gig economy' describes workers who are employed on a short-term or freelance basis. What is the current Labour Party's policy on these types of workers? & There is no specific stance from the Labour Party on gig economy workers, reflecting broader uncertainties and the inherent flexibility and challenges of gig work. \\
\hline
What is the difference between a gig economy and a hustle economy? & 
The gig economy involves short-term, freelance work found via apps or platforms, contrasting with the hustle economy's focus on diverse income streams from side hustles and entrepreneurial activities. The former emphasizes work structure, while the latter covers a wider range of income-generating pursuits, reflecting a shift towards flexible and autonomous earning methods. \\
\hline

\caption{\label{quora-table}First 10 questions and answers obtained from Quora for "gig economy"}
\end{longtable}

Table \ref{chatgpt-table} shows the first 10 questions obtained from the equivalent ChatGPT summary of the search.

\begin{longtable}{*{11}{|p{5.6cm}}|}%{|1|1|}
\hline \multicolumn{1}{|c|}{\textbf{Question}} & \multicolumn{1}{c|}{\textbf{Summary of Answer}} \\ \hline 
\endhead
\endfoot

What is the gig economy and how does it work? & The gig economy is a labor market characterized by the prevalence of short-term contracts or freelance work as opposed to permanent jobs. It allows people to work flexibly, often through apps or online platforms that connect them with customers for short-term services or tasks. \\
\hline
How has the gig economy changed employment? & The gig economy has significantly altered the employment landscape by providing flexible work options, changing the traditional employer-employee relationship, and increasing the number of independent contractors. However, it has also raised concerns about job security and benefits. \\
\hline
What are the pros and cons of the gig economy? & Pros include flexibility, autonomy, and the ability to work from anywhere. Cons involve lack of job security, benefits, and potential for lower income compared to full-time employment. \\
\hline
Is the gig economy sustainable in the long term? & Opinions vary, but the top answer suggests that while the gig economy offers flexibility and opportunities for income, its sustainability is challenged by issues related to workers' rights, benefits, and the need for regulatory frameworks to protect gig workers. \\
\hline
How does the gig economy affect workers' rights? & The gig economy has been criticized for undermining workers' rights, as many gig workers are classified as independent contractors, which exempts them from minimum wage laws, overtime pay, and other benefits that employees typically receive. \\
\hline
What role does technology play in the gig economy? & Technology is the backbone of the gig economy, providing the platforms that connect workers with tasks. It enables the flexibility and accessibility that characterize gig work but also raises questions about data privacy and the digital divide. \\
\hline
Can the gig economy replace traditional employment? & The top answer argues that while the gig economy offers alternatives to traditional employment, it is unlikely to replace it entirely due to the need for stability and benefits that traditional jobs provide. The gig economy complements rather than replaces traditional employment. \\
\hline
What are the biggest challenges facing the gig economy? & Challenges include ensuring fair wages, providing benefits and job security for workers, and creating regulatory frameworks that protect both workers and consumers in the rapidly evolving labor market. \\
\hline
How do gig economy platforms make money? & Gig economy platforms typically earn money by taking a percentage of the transaction or service fee from each gig completed through their platform. This model varies by platform and service type. \\
\hline
What is the future of the gig economy? & The future of the gig economy is seen as evolving, with potential for growth and increased regulation to protect workers. It may see more integration with traditional employment models and increased recognition of gig workers' rights. \\
\hline

\caption{\label{chatgpt-table}First 10 questions and answers obtained from a ChatGPT simulation of a Quora search for "gig economy"}
\end{longtable}

ChatGPT was used to develop a way to categorize the questions. The following eight categories were determined.

    \begin{addmargin}[3em]{3em}%
    \textbf{Definition and Structure}\\
        This category includes questions that define the gig economy and differentiate it from other economic models, such as the digital economy or hustle economy. It explores the structural aspects and characteristics of gig work.
    
    \textbf{Economic Impact}\\
        Questions under this category analyze the economic implications of the gig economy, including its impact on traditional employment, income opportunities, and the broader economy.
    
    \textbf{Workforce Dynamics}\\
        This category focuses on the changes in workforce dynamics due to the gig economy, such as employment patterns, job security, and the relationship between employers and gig workers.
    
    \textbf{Pros and Cons}\\
        Questions that discuss the advantages and disadvantages of participating in the gig economy, from both the workers' and employers' perspectives, are categorized here.
    
    \textbf{Legislation and Regulation}\\
        This category includes questions related to the legal and regulatory frameworks that could or do impact the gig economy, focusing on workers' rights, job security, and platform accountability.
    
    \textbf{Technological Impact}\\
        Questions in this category explore the role of technology in enabling and shaping the gig economy, including discussions on platforms, digital divide, and privacy concerns.
    
    \textbf{Sustainability and Future Outlook}\\
        This category examines the long-term viability and future prospects of the gig economy, considering trends, potential challenges, and opportunities for evolution.
    
    \textbf{Comparative Analysis}\\
        Questions that compare the gig economy with traditional employment or other forms of non-traditional work (e.g., hustle economy) are included in this category, focusing on distinguishing features and implications.
    \end{addmargin}

Table \ref{quora-cat-table} maps the Quora questions against the identified categories.

\begin{longtable}{|p{6cm}|c|c|c|c|c|c|c|c|}
\hline
\textbf{Question} & \rotatebox{90}{\textbf{Definition and Structure}} & \rotatebox{90}{\textbf{Economic Impact}} & \rotatebox{90}{\textbf{Workforce Dynamics}} & \rotatebox{90}{\textbf{Pros and Cons}} & \rotatebox{90}{\textbf{Legislation and Regulation}} & \rotatebox{90}{\textbf{Technological Impact}} & \rotatebox{90}{\textbf{Sustainability and Future Outlook}} & \rotatebox{90}{\textbf{Comparative Analysis}} \\ \hline
What is the gig economy, and how can one participate in it? & x &  &  &  &  &  &  &  \\ \hline
How does the gig economy impact the way people earn money in today's workforce? &  & x & x &  &  &  &  &  \\ \hline
Have you explored the gig economy or sharing economy for income opportunities? &  & x & x &  &  &  &  &  \\ \hline
What are the pros and cons of pursuing a career in the gig economy? &  &  &  & x &  &  &  &  \\ \hline
What are some profitable opportunities in the gig economy in 2024? &  & x &  &  &  &  & x &  \\ \hline
What is the gig economy? What is the digital economy? & x &  &  &  &  & x &  & x \\ \hline
What impact has the gig economy had on recent events? &  & x &  &  &  &  &  &  \\ \hline
How can the gig economy be regulated to protect workers' rights and job security? &  &  &  &  & x &  &  &  \\ \hline
The 'gig economy' describes workers who are employed on a short-term or freelance basis. What is the current Labour Party's policy on these types of workers? &  &  & x &  & x &  &  &  \\ \hline
What is the difference between a gig economy and a hustle economy? & x &  &  &  &  &  &  & x \\ \hline
\caption{\label{quora-cat-table}Mapping of Quora questions to categories}
\end{longtable}

Table \ref{chatgpt-cat-table} maps the ChatGPT generated versions of Quora questions and answers against the identified categories.

\begin{longtable}{|p{6cm}|c|c|c|c|c|c|c|c|}
\hline
\textbf{Question} & \rotatebox{90}{\textbf{Definition and Structure}} & \rotatebox{90}{\textbf{Economic Impact}} & \rotatebox{90}{\textbf{Workforce Dynamics}} & \rotatebox{90}{\textbf{Pros and Cons}} & \rotatebox{90}{\textbf{Legislation and Regulation}} & \rotatebox{90}{\textbf{Technological Impact}} & \rotatebox{90}{\textbf{Sustainability and Future Outlook}} & \rotatebox{90}{\textbf{Comparative Analysis}} \\ \hline
What is the gig economy and how does it work? & x &  &  &  &  & x &  &  \\ \hline
How has the gig economy changed employment? &  & x & x &  &  &  &  &  \\ \hline
What are the pros and cons of the gig economy? &  &  &  & x &  &  &  &  \\ \hline
Is the gig economy sustainable in the long term? &  &  &  &  &  &  & x &  \\ \hline
How does the gig economy affect workers' rights? &  &  & x &  & x &  &  &  \\ \hline
What role does technology play in the gig economy? &  &  &  &  &  & x &  &  \\ \hline
Can the gig economy replace traditional employment? &  &  & x &  &  &  &  & x \\ \hline
What are the biggest challenges facing the gig economy? &  &  & x &  & x &  & x &  \\ \hline
How do gig economy platforms make money? &  & x &  &  &  & x &  &  \\ \hline
What is the future of the gig economy? &  &  &  &  & x &  & x &  \\ \hline
\caption{\label{chatgpt-cat-table}Mapping of ChatGPT questions to categories}
\end{longtable}

\begin{figure}
    \centering
    \includegraphics[width=0.95\linewidth]{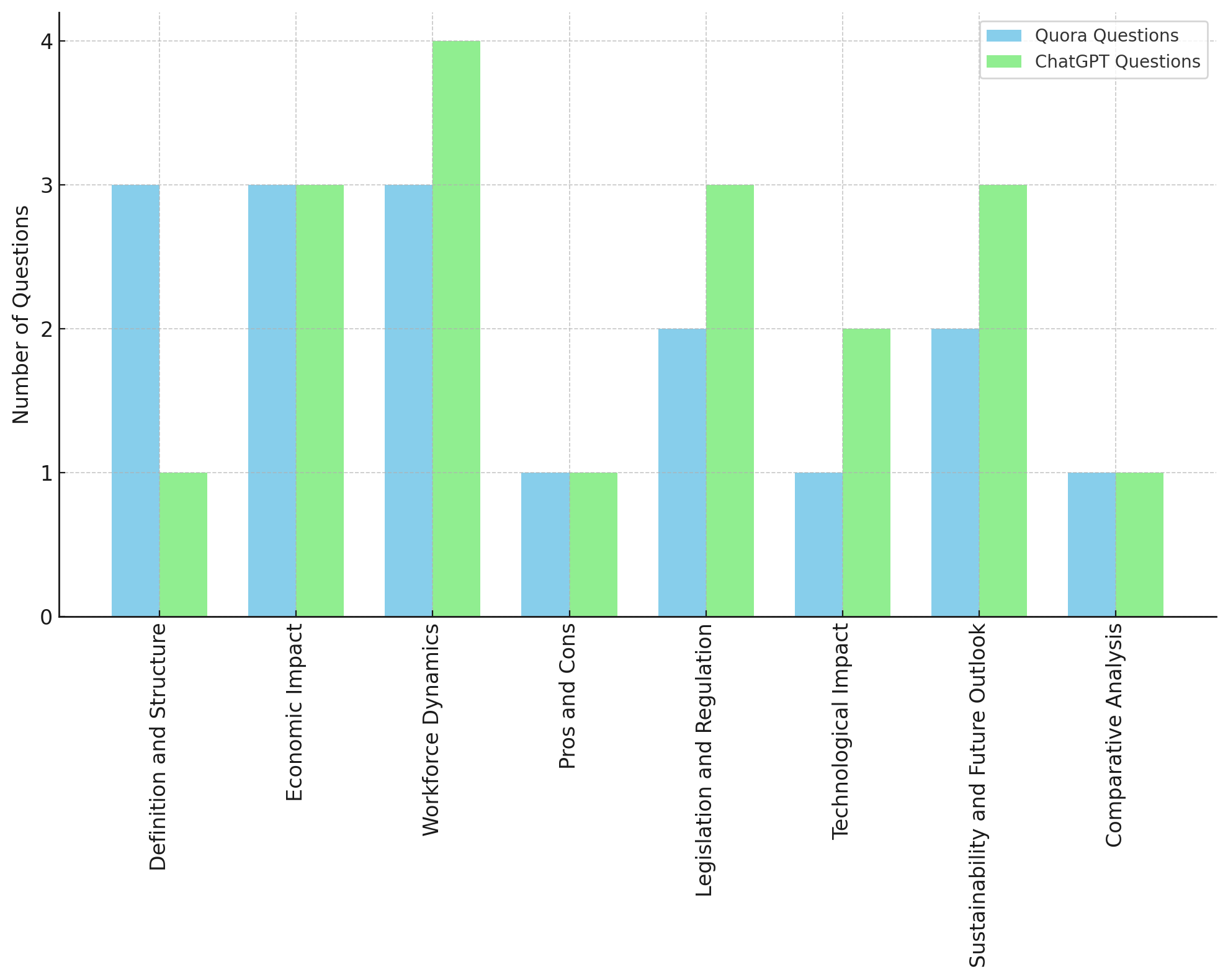}
    \caption{Comparison of categories for Quora and ChatGPT questions}
    \label{fig:cat-comp-fig}
\end{figure}

Figure \ref{fig:cat-comp-fig} shows a comparison between the categories of Quora questions and ChatGPT simulated Quora questions.

\section{Discussion}

The comparison between Quora questions and answers and those simulated by ChatGPT on the topic of the gig economy reveals distinct differences in the focus and depth of content. Quora questions tend to be more definition-focused, often asking "What is the gig economy?" and seeking detailed explanations of its mechanisms and implications. In contrast, ChatGPT-generated questions, while still covering foundational aspects of the gig economy, seem to delve more into its impacts, sustainability, and future prospects. This suggests a more analytical approach in the simulation, potentially reflecting ChatGPT's programming to explore broader implications and generate discussion.

Accuracy varies across both platforms. Quora responses seem to benefit from human expertise, offering nuanced insights into the gig economy's effects on industries, employment patterns, and regulatory challenges. However, the accuracy and quality of these answers can vary significantly depending on the expertise of the individual contributors. ChatGPT, on the other hand, provides consistent, if sometimes generic, responses that reflect a broad understanding of the gig economy. Its answers are grounded in the data it was trained on, which may limit its ability to provide the most current insights or address very recent developments.

Quora's detailed and often personalised answers contrast with ChatGPT's more generalised responses. However, ChatGPT's ability to simulate a wide range of questions and answers on demand and without the variability of human contribution quality is noteworthy. The simulation accuracy for ChatGPT is high in terms of generating relevant and coherent responses, though it may lack the depth and personal experience often found in Quora answers.

Specific analysis related to the identified gig economy categories follow:

    \begin{addmargin}[3em]{3em}%
    \textbf{Definition and Structure}\\
The analysis of Quora and ChatGPT content reveals differences in how each platform approaches the definition and structure of the gig economy. Quora's questions and answers often provide more detailed descriptions and distinctions between the gig economy and related concepts such as the digital and hustle economies. This reflects a deeper exploration into the structural nuances of gig work, offering readers a comprehensive understanding of its characteristics. ChatGPT's responses, while accurate, tend to offer more generalized definitions, focusing on the broad characteristics of gig work without delving as deeply into its distinctions from other economic models.

    \textbf{Economic Impact}\\
When comparing economic impacts, Quora answers typically incorporate personal experiences and specific examples to illustrate the gig economy's effects on traditional employment and the broader economy. These responses often highlight creative destruction, new lifestyle choices, and the challenges of irregular income. ChatGPT, by synthesising a wide range of sources, provides a broader overview of the economic implications, acknowledging both the flexibility offered by gig work and its potential downsides, such as reduced job security. However, ChatGPT might not capture the immediacy and specificity of individual experiences that Quora contributors often provide.

    \textbf{Workforce Dynamics}\\
In discussing workforce dynamics, Quora contributors frequently share firsthand experiences or observations, offering insights into changes in employment patterns, job security, and employer-employee relationships within the gig economy. This personal touch adds depth to the understanding of how gig work affects individuals. ChatGPT, in contrast, outlines these dynamics in a more systematic manner, highlighting general trends and concerns without the personalised narratives found on Quora. This results in a broader, albeit less nuanced, picture of how the gig economy reshapes workforce relationships.

    \textbf{Pros and Cons}\\
The pros and cons of the gig economy are extensively discussed on both platforms. Quora answers often reflect a wide range of opinions, drawing on personal anecdotes to underline the benefits and drawbacks of gig work. This approach provides a rich tapestry of perspectives, showcasing the diversity of experiences within the gig economy. ChatGPT’s responses, synthesising a broad spectrum of viewpoints, present balanced lists of advantages and disadvantages, which, while informative, may lack the depth of personal insight and emotional resonance found in many Quora responses.

    \textbf{Legislation and Regulation}\\
Questions about legislation and regulation receive thoughtful treatment on Quora, where answers sometimes delve into specific legal frameworks, ongoing debates, and policy proposals. These responses can offer detailed insights into the complexities of regulating the gig economy and protecting workers' rights. ChatGPT, with its capacity to draw from a wide array of sources, provides a comprehensive overview of the regulatory challenges and potential solutions but may not always capture the latest developments or nuanced discussions prevalent in some Quora answers.

    \textbf{Technological Impact}\\
The role of technology in enabling and shaping the gig economy is a theme both platforms address with considerable insight. Quora responses often include personal observations on how specific platforms have transformed gig work, touching on issues like data privacy and the digital divide. ChatGPT, leveraging its extensive training data, outlines the fundamental importance of technology to the gig economy, including both its enabling role and the challenges it presents, offering a broad perspective that complements the more specific examples found on Quora.

    \textbf{Sustainability and Future Outlook}\\
Looking at the long-term viability and future of the gig economy, Quora provides a platform for speculative discussions based on personal beliefs, observations, and trends. These discussions can offer diverse viewpoints on the gig economy's sustainability and evolution. ChatGPT's responses, informed by a wide range of sources, present a balanced view of potential future challenges and opportunities, emphasizing the need for regulatory adaptation and highlighting the ongoing debate about the gig economy's role in the future of work.

    \textbf{Comparative Analysis}\\
The comparative analysis between the gig economy and traditional employment or other forms of non-traditional work benefits from the depth and breadth of Quora's user-contributed content, which often includes detailed comparisons based on personal experience or observation. ChatGPT's synthetic approach provides a systematic comparison, outlining key differences and implications without the same level of personal insight or specificity. This highlights how the two platforms can be used in parallel to provide both broad overviews and deep, experience-based insights.
    \end{addmargin}

\section{Conclusion}

This paper has considered how far Quora question and answers can be replicated by the generative AI system ChatGPT. The content analysis has been undertaken with reference to the gig economy, but this method could easily be repeated with reference to other keywords and topics of interest. 

The value of using ChatGPT to replace Quora as a source of information on the gig economy depends very much on the user's needs. For broad, foundational questions and a quick overview of topics, ChatGPT offers a valuable resource that can generate consistent, accurate information efficiently. However, for more detailed, nuanced insights or the latest developments, Quora's platform, which draws on a wide range of human experiences and expertise, referring to platforms such as Uber and AirBnb may be more valuable. ChatGPT's responses, by contrast, appear much more concept-oriented.

The findings underscore the importance of scientific research in assessing these platforms' roles in disseminating knowledge about the gig economy. One limitation of this study is that it has been small in scale. Perhaps a more detailed study would reveal a different level of understanding of these systems. The pace of change in which new generative AI systems are being released also means that any research in this field can quickly become dated.

Continued research can help the research community to understand the strengths and limitations of AI-generated content versus human-contributed content. It can help educators to guide users on when and how to rely on each source effectively. Furthermore, this comparison highlights the need for ongoing evaluation of AI capabilities in accurately simulating human-like responses and the potential for such technology to complement traditional knowledge-sharing platforms.

Ultimately, both Quora and ChatGPT serve valuable but distinct roles in educating the public about the gig economy. While ChatGPT offers a highly accessible, efficient means to obtain information, Quora provides depth, personal insight, and the latest perspectives that only a diverse human community can offer. The complementary nature of these platforms suggests that rather than seeking to replace one with the other, perhaps users and researchers should leverage both to gain a comprehensive understanding of complex topics like the gig economy.

\bibliographystyle{apacite}
\bibliography{references}

\end{document}